\documentstyle[12pt,epsf]{article}

\begin{document}

\begin{titlepage}

\vspace*{2cm}
\begin{center}
{\LARGE  Radiation from Excited Vortex in the
         Abelian  \\  $\;\;\;$ \\
          Higgs Model$^{*}$} \\

\vskip 1cm
by \\
\vskip 1cm
{\large\bf H. Arod\'z and L. Hadasz} \\

\vskip .5cm

Institute of Physics, Jagellonian University,
Cracow$^{**}$
\end{center}

\vskip 1cm

\begin{abstract}
Excitation of a vortex in the Abelian Higgs model is investigated
with the help of a polynomial approximation. The excitation can be
regarded as a longitudinal component of the vector field trapped by
the vortex. The energy and profile of the excitation are found.
Back-reaction of the excitation on the vortex is calculated in the small
$\kappa$ limit. It turns out that in the presence of
the excitation the vortex effectively becomes much wider - its radius
oscillates in time and for all times it is not smaller than the radius
of the unexcited vortex. Moreover, we find that the vector field of the
excited vortex  has long range radiative component.
Bound on the amplitude of the excitation is also found.

\end{abstract}

\vfill

\noindent
\begin{tabular}{l}
TPJU -- 8/95 \\
June 1995    \\
hep-th/9506021
\end{tabular}

\vspace*{2cm}

\noindent
\underline{\hspace*{10cm}}

\noindent
\begin{tabular}{l}
$^{*}$~~Work supported in part by  grant KBN 2 P302 049 05.\\
$ ^{**}$ Address: Reymonta 4, 30--059 Krak\'ow, Poland \\
$\;\;\;$ E--mail: ufarodz@ztc386a.if.uj.edu.pl
\end{tabular}

\end{titlepage}
\baselineskip 16pt

\section{Introduction}

There are many reasons for detailed studies of
dynamics of vortices. Let us mention here the cosmic string hypothesis
\cite{1}, condensed matter physics \cite{2}, QCD flux--tube
\cite{3} and a search for a physical model of a relativistic
string \cite{4}.

Cumulative efforts by various groups of physicists have unravelled
many aspects of fascinating dynamics of vortices. Vortex is a quite
complex object -- due to its macroscopic thread-like extension it
has nontrivial time-evolution with occasional self-interactions
of various kinds like scattering \cite{5} or interconnection
\cite{6}. Moreover, because of its finite transverse size
it can also act as effective potential well trapping various fields.
The first example of this kind was the superconducting string
\cite{7}. It turns out that this can happen also in the more
restrictive framework of Abelian Higgs model \cite{8,9}.
Here the trapped fields are just some components of the
vector field. Because other components of the same vector field
form the vortex itself, it is natural to regard this phenomenon
as the excitation of the vortex. This aspect of dynamics of the vortex
becomes important when the vortex is subjected to such violent transitions
like the interconnection or hard scattering on an external potential.
In fact, numerical investigations of the interconnection clearly show that
after this process the involved parts of vortices are excited, \cite{10}.
The vortex can also become excited by interaction with particles scattered
on it.

Excitations of the vortex in the Abelian Higgs model were studied recently
in the paper \cite{11}. In that paper an effective
 membrane description of the vortex was  proposed. The membrane was
 identified with the surface on which the energy density is maximal - for
 a straight-linear, static vortex with winding number $|n| \geq 2$ it is a
 cylinder of finite radius. In the membrane model there exists solution
which describes a symmetric bulb travelling along the straight-linear
 cylinder. It corresponds to a local excitation of the vortex. The
 shortcoming of such effective model is that it does not yield information
 on detailed form of the  excited vortex fields.

  Comprehensive numerical study of various excitations of local and global
   vortices has been carried out by the authors of Ref.\cite{12}.
The excitation we consider in the present  paper has not been investigated
 there.

 In general, excitations of the vortex can be divided into two classes,
according to the involved components of fields. In the first class there
are excitations which do not involve new components of the gauge
or Higgs fields --  only modes of those components which form
the vortex itself are excited. For the vortex these are the modulus of the
Higgs field and the $A_{\theta}$ component of the vector field ($\theta$
denotes the azimuthal angle). The other class of excitations requires to
extend the vortex Ansatz by allowing for non-trivial values of additional
components of the Higgs or the vector fields, i.e. the phase of the Higgs
field or the $A_z$ component of the vector field (the $A_0$ component is
equivalent to the phase by a gauge transformation). These latter excitations
should rather be regarded as bound states of the additional fields with
the vortex.  The excitations are usually investigated in a linear
approximation -- they are treated as small corrections to the basic vortex
fields.

 The excitation we are interested in  belongs to the second class -- it
involves the $A_z$ component. It can be regarded as a bound state of
longitudinal component of the vector field with the vortex.
 In the paper \cite{9} the existence of the excitation
has been proved, and it has been  shown that it can be represented
by a 2-dimensional Proca field which can propagate along the vortex.
These results have been obtained in a linear approximation with
respect to deviations of the fields from the static vortex solution.
Merely a rough description of the excitation was given -- the main
obstacle was the lack of analytic expression for the vortex solution
even in the Bogomol'nyi limit \cite{13}.  Many interesting questions were
left without  answer, e.g., what are properties of the excitation beyond
the linear approximation.

  The excitation seems to exist for all values of the parameters of the
Abelian Higgs model compatible with the existence of the stable vortex of
topological charge +1 \cite{9}. Nevertheless, in the present paper we
concentrate mainly on the  case of very small  $\kappa,$ where it
is possible to give
explicit and relatively simple analytic formulae. Here $\kappa \equiv
m_A/m_H,$ where $m_A$ and $m_H$ denote the masses of the vector and
Higgs particles, respectively.

 We would like to present a  detailed study of the excitation in the case
$\kappa <1/2$. Our considerations are based on
approximate, analytic expression for the vortex functions obtained in
(appropriately adapted) so called polynomial approximation, used
till now for domain walls \cite{14}. Next, we obtain formulae
 for the profile and energy of the excitation. This in turn makes it
possible to calculate back-reaction of the excitation on the vortex.
The back-reaction is
calculated in the case of very small $\kappa$ when certain simplifications
 occur. We find that in the presence
of the excitation the vortex has oscillating radius, and we calculate the
 frequency and amplitude of these oscillations.  We also notice that there
seems to exist an upper bound on amplitude of the excitation. Such a bound
 reflects the non--linear character of the vortex dynamics and it could
 not be found in  the linear approximation.  However, the most striking
 result is that the excited vortex field has a long range component which
 is interpreted as radiation of the vector field. Also this phenomenon is
 absent in the linear approximation.

 The general conclusion of our paper is that the effects of the
 back-reaction are very important -- they qualitatively change the picture
 of the excited vortex obtained in the linear approximation.

The plan of our paper is as follows. In the next Section we present
the polynomial approximation for the vortex. In Section 3 the
profile and energy of the excitation are calculated. In Section 4
we investigate the back-reaction of the excitation on the vortex.
In Section 5 we have collected several remarks about the method used in
 our paper and about the obtained results.

\section{The polynomial approximation for the unexcited vortex}

We consider Euler--Lagrange equations of the Abelian Higgs model:
\begin{equation}
(\partial_\nu+iqA_\nu)(\partial^\nu+iqA^\nu)\Phi + \frac{\lambda}{2}\Phi
(|\Phi|^2-\frac{2 m^2}{\lambda}) = 0,
\end{equation}
\begin{equation}
\partial_\mu F^{\mu\nu} = iq(\Phi^*\partial^\nu\Phi-\Phi\partial^\nu\Phi^*)
 - 2q^2A^\nu|\Phi|^2,
\end{equation}
where $m,q,\lambda > 0.$ The mass of the Higgs particle is equal to
$m_H^2 = 2 m^2,$ while for the vector field $m_A^2 = \kappa^2m_H^2,$
where $\kappa^2 \equiv 2q^2/\lambda.$ Signature of the space-time metric is
(+1,-1,-1,-1).

The solution of Eqs. (1),(2), representing an infinite, straight-linear,
 topological charge +1 vortex lying along the $x^3$--axis, is sought with
the help of the axially symmetric Ansatz
\begin{equation}
\Phi = \sqrt{\frac{2 m^2}{\lambda}}e^{i\theta}F(\rho),
\end{equation}
\begin{equation}
A_0 = A_3 = 0,
\end{equation}
\begin{equation}
A^1 = \frac{x^2}{\rho}H(\rho), \;\; A^2 = - \frac{x^1}{\rho}H(\rho),
\end{equation}
where $\rho = \sqrt{(x^1)^2+(x^2)^2}$ is the radius in the $(x^1,x^2)$
plane and $\theta = \arctan(x^2/x^1)$ is the azimuthal angle. The functions
$F,H$ obey the following boundary conditions:
\begin{equation}
F(0) = H(0) = 0,
\end{equation}
\begin{equation}
F(\infty) = 1, \;\; \lim_{\rho\to\infty}\rho H(\rho) = -\frac{1}{q}.
\end{equation}
     From (1-5) we obtain equations for $F$ and for
$\chi \equiv q\rho H(\rho) + 1$:

\begin{eqnarray}
F'' + \frac{F'}{r} - \frac{1}{r^2}\chi^2F + \frac12(F-F^3) & = & 0, \\
\chi'' - \frac{\chi'}{r} - \kappa^2 F^2\chi & = & 0,
\end{eqnarray}
where $r\equiv m_H \rho,$ and $'$ denotes $d/dr.$ Boundary conditions
for $\chi$ have the form
\begin{equation}
\chi(0) = 1, ~~~~ \chi(\infty) = 0.
\end{equation}
The magnetic field of the vortex is $B^i = \delta^{i3}B(r),$ where
\begin{equation}
B = - \frac{m^2_H}{q}\frac{\chi'}{r}.
\end{equation}

The main idea of the polynomial approximation is to approximate the fields
 inside the axially symmetric vortex by simple polynomials in $r,$ and
  to match the polynomials
smoothly with asymptotic form of the vortex  fields found for $r>r_0$, where
$r_0$ is the matching point. We shall call $r_0$ the matching radius of the
vortex. Accuracy of such approximation in general increases with the
order of the polynomials. In the present Section we shall assume a third
order polynomial for $F$ and a sixth order polynomial for $\chi.$
Inserting these polynomials in Eqs. (8),(9) we find that the
polynomial for $F$ contains only odd powers of $r$, and the polynomial  for
$\chi$ only even ones. Thus, denoting these polynomials by
underlined letters, we have
\begin{eqnarray}
{\underline F} & = & f_1r - \frac{1}{3!}f_3r^3, \\
{\underline\chi} & = & 1 -
\frac12h_2r^2 + \frac{1}{4!}h_4r^4 - \frac{1}{6!}h_6r^6.
\end{eqnarray}

The order of the
polynomial $\underline\chi$ is the maximal one compatible with
the assumed third order for the polynomial $\underline{F}$ --
in order to compute the term $\sim r^8$ in $\underline\chi$ we would have
 to know the function $F$
in Eq. (9) up to $r^5.$ Equations (8), (9) give the
following recurrence relations for the coefficients of the polynomials
\begin{eqnarray}
f_3 & = & \frac34(\frac12+h_2)f_1, \nonumber \\
h_4 & = & 3\kappa^2f_1^2,  \\
h_6 & = & 5\kappa^2f_1(2f_3+3h_2f_1) \nonumber
\end{eqnarray}

The asymptotic values $\tilde{F},\tilde{\chi}$ of the fields $F,\chi$ outside
 of the vortex, i.e. for
$  r > r_0$, where  $r_0$ will be determined later on, we find
 using  Eqs.(8),(9) and the boundary conditions (7),(10). There is a
 complication due to the fact that the asymptotics of the Higgs field
 depends on $\kappa$ \cite{17}. For $\kappa < 1/2$ it is governed by the
 $\chi^2 F/r^2$ term in Eq.(8), while for $\kappa \geq 1/2$ it is governed
 by the Higgs field mass term. In the following we shall assume that
 $\kappa <1/2$. This is mainly to limit the body of our paper --
 our methods of calculations of the properties of the excitation can be
 applied for any $\kappa$. An additional reason is that for small $\kappa $
 certain simplifications are possible, e.g., one can use a kind of adiabatic
 approximation.

      If for  the Higgs field we take
 \begin{equation}
  \tilde{F}^{(0)} = 1,
  \end{equation}
then in the asymptotic region Eq.(9) is effectively simplified to
\[
\tilde{\chi}'' - \frac{\tilde{\chi}'}{r} - \kappa^2\tilde{\chi} = 0,
\]
and its vanishing for large $r$ solution is given by
\begin{equation}
\tilde{\chi}^{(0)} = c_0 r K_1(\kappa r),
\end{equation}
where $K_1$ is the modified Hankel function \cite{15} and $c_0$
is a constant (to be determined later on). In the next step, one could
insert the asymptotics
(16) for $\chi$ into Eq.(8) and to determine corrections
to the leading asymptotics (15). Iterating these steps and taking
higher order polynomials for $\underline{F}$ and $\underline{\chi}$ we
can obtain a very accurate
description of the vortex which will be presented in a separate
paper \cite{16}. In the present paper, devoted to the excitation
 rather than to the vortex, we
will be satisfied with rather crude description of the unexcited vortex.

The last step in obtaining the static vortex solution is to match
the asymptotics with the polynomials (12), (13).
We know from \cite{17} that for the exact static vortex
solution $F$ is real analytic in $(x^1,x^2).$ As for the magnetic field,
it is also true at least in the Bogomol'nyi limit. For this
reason, the  number of the matching relations depends only on the number of
arbitrary constants present in the asymptotic solutions and in the
polynomials - we just require  maximal
possible smoothness of the fields within the adopted Ansatz (12,13,15,16).
In the present paper we will
approximate $F$ and $\chi$ by functions of $r$ of the
$C^1$ class at $r=r_0$, and real analytic for $r \neq r_0$.
More accurate and smoother approximations to the vortex
 fields will be
presented in \cite{16}. Thus, we choose the following matching conditions
for $F$ and $\chi$:
\begin{equation}
{\underline F}(r_0-) = \tilde{F}(r_0+),
   \;\; {\underline F}'(r_0-) = \tilde{F}'(r_0+),
\end{equation}
\begin{equation}
\underline\chi(r_0-) = \tilde\chi(r_0+), \;\;
\underline\chi'(r_0-) = \tilde\chi'(r_0+).
\end{equation}
Here $g(r_0\pm)\equiv\lim_{r\to r_0\pm}g(r)$ with $g = F,\chi.$
The radius $r_0$ and the coefficients $c_0,f_1,h_2$ are to be determined
from (17),(18). Conditions (17) with (15) taken into acount give
\begin{equation}
f_1 = \frac{3}{2r_0},  \;\; f_3 = \frac{3}{r_0^3}.
\end{equation}
i.e.
\begin{equation}
\underline{F}^{(0)} = \frac32\frac{r}{r_0}
- \frac12\left(\frac{r}{r_0}\right)^3.
\end{equation}
Next, we calculate $h_2, h_4,$ and $h_6$ from formulae (14)
-- in this way we find
\begin{equation}
\underline{\chi}^{(0)} = 1 + \left(\frac14-\frac{4}{3r_0^2}\right)r^2 +
           \frac{9}{32}\frac{\kappa^2}{r_0^2}r^4 +
\frac{3}{16}\frac{\kappa^2}{r_0^2}\left(\frac18-\frac{1}{r_0^2}\right)r^6.
\end{equation}

The final step is to obey the matching
conditions (18) for $\chi.$  Taking $\tilde{\chi}=\tilde{\chi}^{(0)}$ we
 find that they are equivalent to the
following two equations from which we can determine $c_0$ and $r_0$
\begin{eqnarray}
-\frac13 +\frac14(1+\frac{3}{8}\kappa^2)r_0^2 +
\frac{3}{128}\kappa^2r_0^4 & = &
c_0r_0K_1(\kappa r_0), \\
 & & \nonumber \\
\frac{8 - \frac32r_0^2-
\frac{27}{64}\kappa^2r_0^4}{1-\frac{3}{4}(1+\frac{3}{8}\kappa^2)r_0^2 -
\frac{9}{128}\kappa^2r_0^4} & = &
1 + \kappa r_0\frac{K_1'(\kappa r_0)}{K_1(\kappa r_0)}.
\end{eqnarray}
Equation (23) can be easily solved for $r_0$ by numerical methods
for any given $\kappa.$ For very small $\kappa$ one can also
give analytic formula:
\begin{equation}
r_0^2\cong\frac{16}{3}(1+\kappa^2\ln \kappa) +{\cal O}(\kappa^2).
\end{equation}
(we have used the formula $K_1(z) \cong 1/z$ for $0 < z << 1$).
Thus, for $\kappa\to0$ the Higgs radius of the vortex reaches the fixed value
$4/\sqrt{3}\approx 2.31$. Comparison with numerical solutions of Eq.(23)
shows that formula (24) gives slightly too big value of $r_0$, and that the
error is only 6\% even for $\kappa = 0.3$. For $\kappa =0.1$ the error is
1\% and it decreases rapidly for smaller $\kappa$.  In general, the radius
decreases with increasing   $\kappa$, see Fig.1.

              $\;\;$~~~~~ \epsfbox{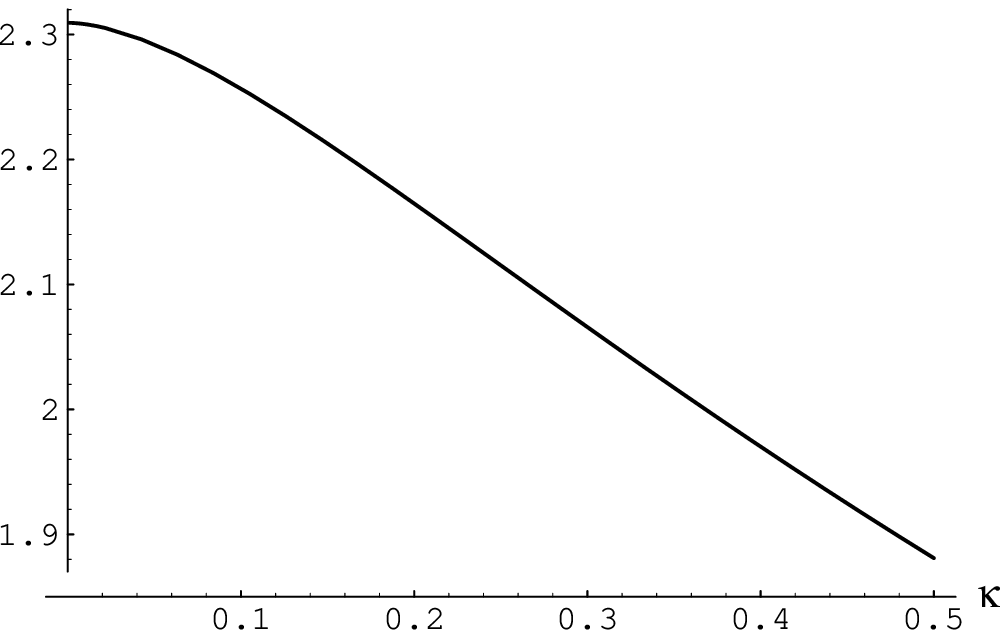}

\centerline{Fig.1. $r_0$ as a function of $\kappa$, determined from Eq.(23).}
\vskip 0.5cm

The constant $c_0$ is found from Eq(22). In particular, for
 $\kappa \rightarrow 0 $
\begin{equation}
c_0 \cong \kappa  (1 - \frac{4}{\sqrt{3}} \kappa^2 \ln \kappa)
 +{\cal O}(\kappa^3).
\end{equation}

The magnetic field is given on outside of the vortex by
formula (11) with $\chi=\tilde{\chi}^{(0)}$ (formula (16). It
has indefinitely increasing range in the limit $\kappa \rightarrow 0$. At
the same time its magnitude decreases so that its integral gives the unit of
the magnetic flux.

Let us  remark that for $\kappa < 1/2$ a more accurate
approximation for the asymptotic field $F$ is provided by the formula
\[ \tilde{F} = \sqrt{ 1 - \frac{2}{r^2} \tilde{\chi}^{(0)2} }. \]
Using this asymptotics instead of $\tilde{F}^{(0)}$ in the matching
conditions (17),(18) and  a fifth order polynomial for $\underline{F}$
we find that for $\kappa \to 0$ $r_0 \cong 2.11$. Thus, the
improved asymptotics gives slightly smaller matching radius.
Such more accurate asymptotics $\tilde{F}$ and a fifth order polynomial
$\underline{F}$ are used in
Section 4. Yet more accurate asymptotics is given by the formula
\[ \tilde{F} = \sqrt{ 1 - \frac{2}{r^2} \tilde{\chi}^{(0)2} } + c K_0(r), \]
where $c$ is a constant. This asymptotics is considered in \cite{16}.

 The approximate, analytic form of the vortex functions $F, \chi$ we have
obtained in this Section for $\kappa < 1/2$  will be used to study the
excitation in more detail than it was done in \cite{9}.

\section{Analytic description of the excitation in  linear approximation}

The excitation is another vortex--like  solution of Eqs. (1,2). It
belongs to  the same topological class as the vortex itself.
Arguments for existence of such a solution, based on a linear approximation,
are given in \cite{8},\cite{9}. This solution is time-dependent.

The Ansatz (3-5) is now extended by including
$A^3 \equiv A(t,r)$, while $A_0 = 0$ again. The $F$ and $\chi$
functions are now time--dependent. In this case equations (1),(2) reduce
 to the following set of equations
\begin{eqnarray}
-\ddot{F} + F'' + \frac{F'}{r} - \left(\frac{1}{r^2}\chi^2
+\frac{q^2}{m_H^2}A^2\right)F + \frac12(F-F^3) & = & 0, \\
-\ddot\chi + \chi'' -\frac{\chi'}{r} - \kappa^2F^2\chi & = & 0, \\
-\ddot{A} + A'' + \frac{A'}{r} - \kappa^2F^2A & = & 0,
\end{eqnarray}
where $ t\equiv  m_H x^0$ and the dots denote $d/dt.$
Notice that there is no
direct coupling of the $A$ field to the $\chi$ field, and that
$A$ influences $F$ only by the single term $\frac{q^2}{m_H^2}A^2F$
in Eq.(26).

We shall seek for the solutions of
this set of equations using an iterative procedure. First, in the present
Section we aproximately solve Eq.(28) assuming that the fields $F$ and
$\chi$  form a background not influenced by the
$A$ field, i.e. that they are equal to the initial vortex fields
discussed in the previous Section. With the back-reaction of the $A$ field
on the vortex switched off, equation (28) becomes linear equation for $A$
with explicitely given function $F(r)$ -- hence the name "linear
 approximation". The back-reaction of the $A$  field on the $F$ and $\chi$
 fields we shall calculate in the next Section.  We hope that if we
continued the iterations we would obtain a sequence of approximations
convergent towards the exact solution at least for sufficiently small
$\frac{q^2}{m_H^2}A^2$ (which is a dimensionless quantity).

Thus, in the present Section we will find the bound-state type solution
of the equation
\begin{equation}
- \ddot{A} + A'' +\frac{A'}{r} - \kappa^2 F^{(0)2} A = 0,
\end{equation}
where $F^{(0)}$ denotes the Higgs field of the unexcited vortex, calculated
in the previous Section.
 We shall use the crude asymptotics
(15) and the third order polynomial (12) for the Higgs field -- it is
sufficient in order to find the excitation and its basic characteristics.
More accurate form of the Higgs field would  require significantly longer
 calculations and we think that the improvement would yield merely
more accurate numbers, with no change in the overall picture of the excited
vortex.
If $r\leq r_0$, the functions $F,\chi$ and $A$ are again approximated
by polynomials
$\underline{F},\underline\chi$ and $\underline{A},$ respectively. For
$\underline{F}$ and  $\underline\chi$ we have formulae (12,13), and
\begin{equation}
\underline{A} = a_0(t) - \frac12a_2(t)r^2 + \frac{1}{4!}a_4(t)r^4 -
         \frac{1}{6!}a_6(t)r^6.
\end{equation}
We have  noted that the coefficients in $\underline{A}$ can be
time--dependent. Inserting formula (30) in
Eq.(28) we obtain the following set of equations
\begin{eqnarray}
\ddot{a}_0 & = & -2a_2, \\
\ddot{a}_2 & = & -\frac{4}{3}a_4 + 2\kappa^2a_0f_1^2, \\
\ddot{a}_4 & = & -\frac{6}{5}a_6 + 8\kappa^2a_0f_1f_3 + 12\kappa^2a_2f_1^2.
\end{eqnarray}

If $r\!>\!r_0$, in Eq.(29) we put $F^{(0)}=1$. It follows that
$A=\tilde{A},$ where
\begin{equation}
\tilde{A} = c_1\cos(\omega_0t + \delta)K_0(k_0r).
\end{equation}
Here $c_1$ is a constant, $K_0$ is the modified Hankel function and
\begin{equation}
\omega_0^2 +k_0^2 = \kappa^2.
\end{equation}

The frequency $\omega_0$ and the constant $c_0$ are determined from
matching conditions at $r = r_0:$
\begin{eqnarray}
\underline{A}(r_0-) & = & \tilde{A}(r_0+), \\
\underline{A}'(r_0-) & = & \tilde{A}'(r_0+).
\end{eqnarray}
We assume that the coefficients $a_{2k}, k=0,1,2,3,$ have the same
time--dependence as $\tilde{A},$ i.e.
\[
a_{2k} = \alpha_{2k}\cos(\omega_0t+\delta)
\]
with constant $\alpha_{2k}$. Then, using Eqs.(31-33), (19) we can express
$\alpha_2,\alpha_4,\alpha_6$ by $\alpha_0,r_0,\kappa.$  We obtain that
\begin{equation}
\underline{A} =  \underline{\alpha}(r)\cos(\omega_0 t + \delta),
\end{equation}
where
\begin{equation}
\underline{\alpha}(r) = \alpha_0 \left[1-\frac14\omega_0^2r^2 +
         \frac{1}{64}\left(\omega_0^4+\frac{9}{r_0^2}\kappa^2\right)r^4 -
         \frac{1}{48}\left(\frac{45}{8}\frac{\omega_0^2\kappa^2}{r_0^2} +
         12\frac{\kappa^2}{r_0^4} + \frac18\omega_0^6\right)r^6\right].
\end{equation}
The matching conditions (36),(37) are equivalent to the
following two equations
\begin{equation}
1 - \frac14\omega_0^2r_0^2 + \frac{1}{64}\omega_0^4r_0^4 -
\frac{1}{384}\omega_0^6r_0^6 - \frac{7}{64}\kappa^2r_0^2 -
\frac{15}{128}\omega_0^2\kappa^2r_0^4 =
      \frac{c_1}{\alpha_0}K_0(k_0r_0),
\end{equation}

\begin{equation}
\frac{\omega_0^2r_0^2 - \frac18\omega_0^4r_0^4 + \frac{1}{32}\omega_0^6r_0^6
 +   \frac{45}{32}\omega_0^2\kappa^2r_0^4 + \frac{15}{8}\kappa^2r_0^2}
  {1 - \frac14\omega_0^2r_0^2 +\frac{1}{64}\omega_0^4r_0^4 -
   \frac{1}{384}\omega_0^6r_0^6 - \frac{7}{64}\kappa^2r_0^2 -
   \frac{15}{128}\omega_0^2\kappa^2r_0^4} =
       -2r_0k_0\frac{K_0'(k_0r_0)}{K_0(k_0r_0)},
\end{equation}
where $K_0'(z)\equiv\frac{d}{dz}K_0(z)$ with $z=k_0r_0,$ and $r_0$
is to be calculated from Eq.(23) of the previous Section.

      From Eq.(40) we can determine only the ratio $c_1/\alpha_0,$
so the amplitude $\alpha_0$ remains arbitrary in accordance with
linearity of Eq.(29) with the fixed $F^{(0)}.$

Equation (41) gives $\omega_0$ as a function of $\kappa.$ In
general, one has to use numerical methods to find this function.
It is plotted in Fig.2.

$ \;\;\;$   \epsfbox{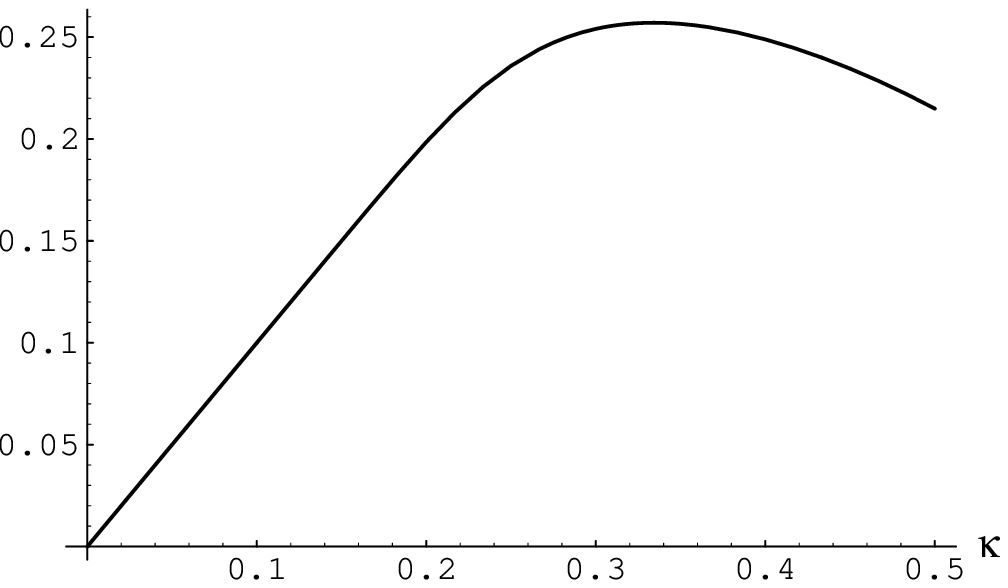}

 \centerline{Fig.2. $\omega_0$ as the function of $\kappa$, determined
            from Eqs.(41),(35)}.

 If $\kappa\ll1$, one can use formula (24) and the following asymptotic
formula for the modified Hankel function \cite{15} for $0 < z << 1$
\[
\frac{zK_0'(z)}{K_0(z)} \cong \frac{1}{\ln(\gamma z) - \ln2 },
\]
where $\gamma\approx1.781072$ is the Euler constant.
Formula (35) implies that for $\kappa\to0 ~~\omega_0$ and
$k_0$ vanish. Keeping only the leading terms on both sides of
Eq.(41) we obtain the equation
\[
(-\frac{23}{16}\kappa^2 + \frac12 k_0^2) r_0^2 \approx \frac{1}{\log k_0},
\]
which gives
\[
k_0^2 \sim \exp\left\{-\frac{32}{23}
\frac{1}{\kappa^2}\frac{1}{r_0^2}\right\}
\]
It is easy to see that the value of the coefficient in front of
this formula for $k_0$ depends on subleading terms
in Eq.(41). Taking these terms into account we find that for
$\kappa\ll1$
\begin{equation}
k_0^2 \cong 3.86\frac{1}{r_0^2}
\exp\left\{-\frac{32}{23}\frac{1}{\kappa^2}\frac{1}{r_0^2}\right\},
\end{equation}
where $r_0$ is given by formula (24).
Comparison with numerical solutions of Eq.(41) shows that formula (42)
gives slightly too big value of $k_0$, with 13\% accuracy even for
$\kappa =0.3$. For $\kappa = 0.1$ the error is 2\% and it decreases rapidly
 for smaller  $\kappa$.

{}From Eq.(40) we find that for small $\kappa$
\begin{equation}
\frac{c_1}{\alpha_0} \cong - \frac{23}{16}r_0^2\kappa^2.
\end{equation}
It follows that when $\kappa \rightarrow 0$, the function
$c_1 K_0(k_0 r) $ for any finite fixed $r$ tends to a constant value
equal to $\alpha_0$, hence $\tilde{A} \rightarrow a_0$.

 The existence of the excitation in the linear approximation can be seen
  by regarding the equation
\[
-\frac12\left(\alpha''+\frac{\alpha}{r}\right) +
\frac{\kappa^2}{2}F^{(0)2}\alpha = \frac{\omega_0^2}{2}\alpha,
\]
obtained from Eq. (29) by substituting
$A = \cos(\omega_0t+\delta)\alpha(r),$ as  Schr\"odinger equation for
a particle in the 2--dimensional potential well given by
$V(r) = \frac{\kappa^2}{2}F^{(0)2}.$
It is a well known fact \cite{18} that any potential well
on a plane has at least one bound state, hence localized non--zero
solution $A$ of Eq.(29). For small $\kappa$ the potential well
has finite width $r_0\approx4/\sqrt{3}$ and it is very shallow
($V_{\mbox{\scriptsize max}} = \kappa^2/2).$
Nevertheless, the bound state exists and the corresponding frequency
$\omega_0,$ is given by formula (42). Estimate made in \cite{9} shows that
 the potential well is too weak to support p-wave bound states.
Therefore, we do not consider more general, $\theta$-angle dependent
 $A$ field.

\section{The vortex in presence of the excitation}

Mathematically, the presence of the excitation changes the vortex because
of the $A^2$ term in Eq.(26). Because
$A^2 \sim \cos^2(\omega_0t+\delta) = \frac12[1+\cos(2\omega_0t+2\delta)],$
the perturbation has two components: the static one and the one
oscillating with the frequency $2\omega_0.$ In order to calculate their
influence on the vortex we reanalyse  Eqs.(26),(27) assuming that
$A$ is given
by the approximate solution of Eq.(28) found in the previous Section.
 We shall consider the limiting case $\kappa \rightarrow 0 $ when
we can  do this analytically.

 In the case of very small $\kappa$,
$A^2$ changes in time very slowly as the frequency $2 \omega_0$ is
 very small.  Because we
expect that the induced by the presence of the excitation
time dependence of $F$ is also characterized by the
frequency $2\omega_0,$ the term $\ddot{F}$ on the r.h.s of
Eq.(26) is of the order $\omega_0^2 \sim \kappa^2$ and it can be
neglected. Therefore, equation (26) approximately reduces to the following
equation
\begin{equation}
F'' + \frac{1}{r} F' - (\frac{1}{r^2}\chi^2
+ \frac{q^2}{m_H^2} A^2) F
 + \frac{1}{2}(F-F^3)  =  0.
\end{equation}

Now let us turn to the Eq.(27) for the field $\chi$. Again,
 we expect that  the induced time dependence of $\chi$ is characterized by
 the frequency $2\omega_0$, so that the term $\ddot{\chi}$ is of the order
 $\kappa^2$. However, if we neglect this term then we also have to neglect
 the term $\kappa^2 F^2 \chi$, and Eq.(27) reduces to the following equation
 \begin{equation}
 \chi'' - \frac{1}{r} \chi' =0.
 \end{equation}
 Its general solution has the form
 \begin{equation}
 \chi = c_2 + d_2 r^2,
 \end{equation}
 where $c_2, d_2$ are arbitrary constants. It is easy to see that such $\chi$
can not obey the boundary conditions (10). Thus, neglecting the terms
$\sim \kappa^2$ in Eq.(27) altogether is an oversimplification if we consider
the whole range of $r$, i.e. $[0,\infty)$. On the other hand, it
is acceptable approximation
 if we consider $\chi$ in a finite interval of the $r$ variable, e.g., in a
 vicinity of the matching point, now denoted by $r_e$.
It breaks down only if we try to apply it also for $r\rightarrow \infty$.
This means that the limits $\kappa\rightarrow 0$ and $r\rightarrow\infty$
are not interchangeable, and the correct order of them is that first we
should find $\chi$ for very large $r$ and only after that we can
consider the limit $\kappa\rightarrow 0$. One could see this by analysing the
vortex solution of Section 2, cf. formulae (16),(25). Later on in this
 Section
we shall find the asymptotics $\tilde{\chi}$ for the very large $r$.
 For the current calculations it is sufficient to know that for $r\sim r_e$
 $\chi$ is given by formula (46).

{}From Eq.(44) we obtain approximate form of the Higgs field $F$ in the
presence of the excitation. As always in the polynomial approximation we
separately consider the regions of small  and large $r$. The
matching takes place at $r=r_e$.
In the region $r \leq r_e$ we will approximate $F$ by a
 fifth order polynomial,
\begin{equation}
{\underline F}  =  f_1 r - \frac{1}{3!}f_3r^3 + \frac{1}{5!}f_5r^5.
\end{equation}
The reason is that the more precise (from that considered in Section 2)
 form of the asymptotics of the Higgs field which is given
 below requires at least the fifth order polynomial if the matching
conditions (17) are to be satisfied.
Instead of the first of recurrence relations (14) we now have
\begin{equation}
f_3 = \frac{3}{4}\left(\frac12+h_2-\frac{q^2}{m_H^2}a_0^2\right)f_1.
\end{equation}
 We also obtain the relation
\begin{equation}
f_5 = \frac{5}{6}(\frac{1}{2}+h_2-\frac{q^2}{m_H^2}a_0^2)f_3 + \frac{5}{2}
(\frac{1}{6}h_4 +\frac{1}{2}h_2^2 +f_1^2 - \frac{2q^2}{m_H^2}a_0a_2)f_1.
\end{equation}
Notice that passing to the fifth order polynomial does not increase the
number of arbitrary constants in our Ansatz, because the new constant $f_5$
is related to the other constants by formula (49).

 For  $\chi$ inside the vortex we again use the
polynomial  (13), and now the coefficients $h_{2k}$ can in principle depend
on time. However, the corresponding recurrence relations in the limit
$\kappa \rightarrow 0$ reduce to
\begin{equation}
 h_4 \cong 0,\;\; h_6 \cong 0.
 \end{equation}
  The matching conditions
(18) applied to the polynomial $\underline{\chi}$ and $\chi $ given by
formula (46) imply that
\begin{equation}
c_2=1, \;\;\; d_2 = - \frac{1}{2} h_2.
\end{equation}

The asymptotics $\tilde{F}$ of the Higgs field obeys the following equation
\begin{equation}
\tilde{ F}'' + \frac{1}{r}\tilde{ F}' - (\frac{1}{r^2}\tilde{\chi}^2
+ \frac{q^2}{m_H^2} \tilde{A}^2)\tilde{ F}
 + \frac{1}{2}(\tilde{F}-\tilde{F}^3)  =  0,
\end{equation}
where $\tilde{A}$ is given by formula (34), and $\tilde{\chi}$ is found
below. We use the observation following from formulae (34),(43) that
for any $r>0$ in the limit $\kappa \rightarrow 0$
\[ (\tilde{A}^2)' \sim \kappa^2. \]
Hence, $\tilde{A}^2$ changes very slowly with $r$ for small $\kappa$.
As we shall see below,  the same is true for $\tilde{\chi}$, essentially
because the mass of this field is equal to $\kappa$.
Then, it is easy to check that the equation for $\tilde{F}$ has the
following approximate solution
 \begin{equation}
 \tilde{F} = \sqrt{1 -\frac{2}{r^2}\tilde{\chi}^2
  -\frac{2 q^2}{m_H^2}\tilde{A}^2 }.
 \end{equation}
Adopting this solution is equivalent to neglecting the derivative terms
($\sim F', F''$) in Eq.(52).

The next step is to obey the matching conditions (17). Because the matching
takes place at finite $r=r_e$ we may use formula (46). We also use formulae
(50),(51), and we notice that for finite $r$ and $\kappa \rightarrow 0$
\[ \tilde{A}^2 \cong a_0^2, \]
as follows from formulae (34),(42),(43).
The matching conditions (17) give the following relation
\begin{equation}
f_1 =  \frac{8(5\beta r_e^2 -12)}{r_e^2(32-\beta r_e^2) \sqrt{\beta
r_e^2-2}},
\end{equation}
where
\[\beta \equiv 1 - \frac{2 q^2 a_0^2}{m_H^2} + 2 h_2,  \]
as well as the equation
\begin{equation}
x\sqrt{\beta -x} (16x-\beta) =\frac{ 5\beta - 6x}{\sqrt{\beta-x}}
\left(4x^2 - \frac{\beta}{2}x +
\frac{\beta^2}{48} +\frac{8}{3}
\frac{x^3 (5\beta -6x)^2}{\beta-x)(16x-\beta)^2}\right)
\end{equation}
for
\[ x \equiv \frac{2}{r_e^2}. \]
Equation (55) has the following solution
\[ x = d_0\; \beta, \] where the constant $d_0$ can be easily determined by
 numerical methods,
 \[ d_0 \approx 0.449. \]
Moreover, we shall show later on that $h_2 \approx 0$.
Hence,
\begin{equation}
  r_e \approx  \frac{2.11}{\sqrt{1- \frac{2 q^2}{m_H^2}a_0^2 }}.
  \end{equation}

It is clear from this formula  and from the formula
\[ a_0^2 = \alpha_0^2 \cos^2(\omega_0 t + \delta) \]
 that $r_e$ oscillates in the interval $[2.11, 2.11/(1-
  2q^2\alpha^2_0/m_H^2)].$ Let us recall that 2.11 is the matching
 radius in the absence of the excitation, as mentioned at the end
 of Section 2.

We see from formula (56) that as $\alpha_0^2$ approaches $m_H^2/(2q^2)$
the amplitude of oscillations of $r_e$  increases indefinitely,
 and for still higher amplitude
of the excitation formula (56) gives unphysical $r_e$.
We interpret this result as a hint that there is an upper bound
$\sim  m_H/(\sqrt{2}q)$ on the amplitude  $\alpha_0$ of the excitation.
Such a bound would reflect nonlinear character of the set
(26-28) -- in the linear approximation  no restriction on the amplitude
appears. It would be desirable to check the existence of the bound by
another method, e.g., by direct numerical analysis of
Eqs.(26-28). Interesting question what happens to the vortex
if the bound is exceeded we will leave for a separate investigation.

In order to find the $\tilde{\chi}$ field of the excited vortex
 for the very large $r$ we have to
solve equation (27) with $F=\tilde{F}$, where $\tilde{F}$ is given by
formula (53). As discussed earlier, now we should not neglect the terms
 proportional to $\kappa^2$. Inserting formula (53) and neglecting the
 term $2\kappa^2 \tilde{\chi}^3/r^2$ (because we expect that $\tilde{\chi}$
 is small for $r > r_e$ in comparison with the other terms
 -- subsequent computation confirms this expectation)
we obtain the following equation
\begin{equation}
-\ddot{\tilde{\chi}} + \tilde{\chi}'' - \frac{1}{r} \tilde{\chi}'
- m_{eff}^2 \tilde{\chi} + \kappa^2 \frac{q^2c_1^2}{m_H^2} K_0^2(k_0 r)
\cos(2\omega_0t + 2 \delta) \tilde{\chi} =0,
\end{equation}
where we have introduced the notation
\begin{equation}
m_{eff}^2 = \kappa^2 ( 1 - \frac{q^2 c_1^2}{m_H^2} K_0^2(k_0r)).
\end{equation}
Next, we split $\tilde{\chi}$ into a static and time-dependent parts,
\begin{equation}
\tilde{\chi} = \tilde{\chi}_s + \tilde{\chi}_{\omega},
\end{equation}
where the static part by definition obeys the equation
\begin{equation}
 \tilde{\chi}_s^{''} - \frac{1}{r} \tilde{\chi}_s^{'}
- m_{eff}^2 \tilde{\chi}_s =0.
\end{equation}
It follows from formula (59) and Eqs.(57),(60) that
 $\tilde{\chi}_{\omega}$ obeys the equation
\begin{eqnarray}
& -\ddot{\tilde{\chi}}_{\omega} + \tilde{\chi}^{''}_{\omega}
 - \frac{1}{r} \tilde{\chi}^{'}_{\omega}
- m_{eff}^2 \tilde{\chi}_{\omega} +
 \kappa^2 \frac{q^2c_1^2}{m_H^2} K_0^2(k_0 r)
\cos(2\omega_0t + 2 \delta) \tilde{\chi}_{\omega}  & \nonumber \\
& = - \kappa^2 \frac{q^2c_1^2}{m_H^2} K_0^2(k_0 r)
\cos(2\omega_0t + 2 \delta) \tilde{\chi}_{s}.  &
\end{eqnarray}
Thus, $\tilde{\chi}_s$ acts as a source for $\tilde{\chi}_{\omega}$.

In the present paper we will discuss only a perturbative solution of Eq.(61),
obtained by expanding  $\tilde{\chi}_{\omega}$ in powers of the dimensionless
amplitude  $q\alpha_0/m_H$ of the excitation.  This solution is expected to
be a good  approximation to the  exact solution of Eq.(61) if the
 dimensionless amplitude is small, so  from now on we assume that this
  is the case. The perturbative solution is
 sufficient to show that the excited vortex radiates the $\chi$ field.
 We start from the observation that the r.h.s. of Eq.(61) together with
 formula (43)  suggest that  $\tilde{\chi}_{\omega}$ is of the order
$q^2 \alpha_0^2/m_H^2$. Therefore, the last term on the l.h.s. of Eq.(61)
 is of the order $(q^2 \alpha_0^2/m_H^2)^2$ and can be neglected if the
 dimensionless amplitude of the excitation is small. For the same reason,
$m_{eff}$ in Eq.(61) can be replaced by  $\kappa$. Solving obtained
in this way the simplified version of Eq.(61) we find the lowest order
contribution to the perturbative series for $\tilde{\chi}_{\omega}$.

 Let us introduce a new function $h_{\omega}(r)$
\begin{equation}
\tilde{\chi}_{\omega} = \frac{\kappa q^2 \alpha_0^2}{m_H^2}
 \cos(2\omega_0 t + 2 \delta)\; r \; h_{\omega}(r) .
 \end{equation}
 The simplified form of Eq.(61) is equivalent to the following equation
  for the function
 $h_{\omega}$:
 \begin{equation}
 h_{\omega}^{''} +\frac{1}{r} h_{\omega}^{'} - \frac{1}{r^2} h_{\omega}
 + k_1^2  h_{\omega} = -\frac{\kappa}{r}\frac{c_1^2}{
 \alpha_0^2} K_0^2(k_0r) \tilde{\chi}_s (r),
 \end{equation}
 where
 \[   k_1 \equiv \sqrt{4\omega_0^2 - \kappa^2}.   \]
 Observe that for the very small $\kappa$ the coefficient $k_1^2$ is
 positive because then $\omega_0^2 \sim \kappa^2$, see formulae (35),(42).

 The asymptotics $\tilde{\chi}$ is determined by  $\tilde{\chi}_s$
 and $h_{\omega}$. Equation (60) has the following vanishing at the
 infinity, approximate solution
\begin{equation}
\tilde{\chi}_s \approx c_3 \kappa r  K_1(m_{eff}r),
\end{equation}
where $c_3$ is a constant to be determined later.

General, vanishing at the infinity solution of Eq.(63) is given
by the formula
\begin{equation}
h_{\omega}(r) = c_3 h_{\infty}(k_1r) + c_4 J_1(k_1r) + c_5 N_1(k_1r),
\end{equation}
where $c_4,c_5$ are constants to be determined later on; $J_1,N_1$ are Bessel
and  von Neumann  functions, respectively \cite{15};
and $h_{\infty}(z)$ (with $z=k_1r$) is a particular solution of the
inhomogeneous Bessel equation
\begin{equation}
\frac{d^2 h_{\infty}(z)}{dz^2} +\frac{1}{z} \frac{dh_{\infty}(z)}{dz}
+ (1-\frac{1}{z^2}) h_{\infty}(z) = - \frac{\kappa^2}{k_1^2} \frac{c_1^2}{
\alpha_0^2} K_1(\frac{\kappa}{k_1}z) K_0^2(\frac{k_0}{k_1}z).
\end{equation}
For small $\kappa$ we can approximate
 \[ k_1 \approx \kappa \sqrt{3}.  \]
Equation (66) can be solved by a standard method \cite{19},
\begin{equation}
h_{\infty}(z) = f(z) J_1(z) + g(z) N_1(z),
\end{equation}
where
\begin{equation}
f(z) = - \frac{\pi c^2_1}{6\alpha_0^2}
 \int_z^{\infty} dx\; x N_1(x)
K_1(\frac{x}{\sqrt{3}}) K_0^2(\frac{k_0}{\kappa\sqrt{3}}x),
\end{equation}
and
\begin{equation}
g(z) =  \frac{\pi c^2_1}{6\alpha_0^2}
 \int_z^{\infty} dx\; x J_1(x)
K_1(\frac{x}{\sqrt{3}}) K_0^2(\frac{k_0}{\kappa\sqrt{3}}x).
\end{equation}

The constants $c_3,c_5$ can be fixed by matching the asymptotics
$\tilde{\chi}$ with $\chi $, given by formulae (46),(51), at $r=r_e$.
 This requirement gives the following equations
\begin{equation}
\tilde{\chi}_{s}(r_e) + \frac{\kappa q^2 \alpha_0^2}{m_H^2} r_e h_{\omega}
(r_e) \cos(2\omega_0 t + 2 \delta) = 1 - \frac{h_2}{2} r_e^2,
\end{equation}
\begin{equation}
\tilde{\chi}_s^{'}(r_e) + \frac{\kappa q^2 \alpha_0^2}{m_H^2} [
r_e h_{\omega}^{'}( r_e) + h_{\omega}( r_e) ]
  \cos(2\omega_0 t + 2 \delta) =  - h_2 r_e.
\end{equation}
Next, we notice that for small $\kappa$  (when $\kappa r_e <<1$)
\[ \tilde{\chi}_{s}(r_e) \approx c_3, \;\; \tilde{\chi}^{'}_s(r_e)
\approx 0,\]
\[ \kappa r_e J_1(\sqrt{3} \kappa r_e) \approx 0,
 \;\; \kappa r_e N_1(\sqrt{3} \kappa  r_e) \approx
-  \frac{2 }{\sqrt{3} \pi}. \]
Then it is easy to see that conditions (70),(71) imply that
\begin{equation}
 h_2=0, \;\;   c_3 = \sqrt{1 - \frac{q^2 \alpha_0^2}{m_H^2}}, \;\;
   c_5 = - g(0) c_3.
\end{equation}
The definition (69) gives for $z=0$
\begin{equation}
g(0) = \frac{\sqrt{3}\pi}{8}.
\end{equation}

The constant $c_4$ is not fixed by the matching conditions. The
corresponding term in $\tilde{\chi}$ has the following form
\[ \delta \tilde{\chi} = c_4 \frac{\kappa q^2 \alpha^2_0}{m_H^2}
r J_1(\sqrt{3} \kappa r) \cos(2\omega_0t +2 \delta). \]
This function, as well as its derivatives, is regular for all $r$ including
$r=0$, and it obeys the  wave equation
\[ - \delta\ddot{\tilde{\chi}} + \delta \tilde{\chi}^{''} - \frac{1}{r}
\delta\tilde{\chi}^{'} - \kappa^2 \delta\tilde{\chi} =0  \]
in the whole space. Therefore, we shall regard $\delta\tilde{\chi}$ merely as
an artifact of the linearity of Eq.(57), and interpret $\delta\tilde{\chi}$
 as a free, standing electromagnetic wave which is  not related to the
excited vortex.  Because we are interested in the excited vortex alone,
 we put
\begin{equation}
c_4 = 0.
\end{equation}

Thus, we have finally obtained that
\begin{equation}
h_{\omega}(r) = \sqrt{1 - \frac{q^2\alpha_0^2}{m_H^2}}
(h_{\infty}(\sqrt{3}\kappa r) - \frac{\sqrt{3}\pi}{8} N_1(\sqrt{3}\kappa r)).
\end{equation}

 The function $h_{\infty}(z)$ vanishes exponentially for large $z$, but
 von Neumann function $N_1(z)$ decreases rather slowly for large $z$,
 \[ N_1(z) \cong \sqrt{\frac{2}{\pi z}} \sin(z- \frac{3\pi}{4}).  \]
 Thus, the $\tilde{\chi}$ field has the long range component equal to
 \[ - \frac{\sqrt{3}\pi}{8} \frac{q^2\alpha_0^2}{m_H^2}
 (1- \frac{q^2\alpha_0^2}{m_H^2})^{1/2} \kappa r N_1(\sqrt{3}\kappa r)
 \cos(2 \omega_0t + 2 \delta).   \]
 Its  natural interpretation is that it describes radiation of the vector
 field from the excited vortex.

 It is easy to check that the resulting total field
 configuration has infinite
 energy per unit length of the vortex. This is due to the fact that
  we have  considered the
excitation of the standing wave type for the straight-linear, infinite
vortex in infinite space.
The presence of the long range radiative component means that the excited
vortex should not be regarded as a localised soliton, eventhough it has
the  topological charge +1.

\section{Remarks}

 Let us recapitulate results of our paper. First, on a methodological side,
 we have shown that with  the help of the polynomial approximation
 one can obtain analytic description of the excited vortex,
 including the profile and frequency of
 the excitation as well as the back-reaction of the excitation on the
 vortex. Thus, the polynomial approximation turns out to be a useful tool
 also for investigations of dynamics of vortices -- earlier it has been
 applied to domain walls  \cite{14}.

 Second, on a physics side, we have shown that the excited vortex
 contains the  radiative  component. To find it, one has to
  calculate the back-reaction of the excitation on the vortex. The
 back-reaction is due to  non-linearity of Eqs.(26-28). In the linear
  approximation considered in
 Section 3 the effects which are due to the back-reaction are not taken
 into account.  The radiative component we have found is formed of
 the vector field. We expect that after several iterations of our procedure,
 i.e. correcting the vortex fields, then calculating corrections to the
  excitation field $A$, next again  correcting the vortex fields, etc., one
  would also obtain a radiative component in the Higgs field. Its amplitude
  will be proportional to a higher power of the square of the
  dimensionless amplitude $q^2 \alpha_0^2/m_H^2$ of the excitation.
Taking the back-reaction into account has resulted in drastic alteration
of physical characteristics of the excited vortex obtained in the linear
approximation.

In the present paper we have considered for simplicity only the standing
 wave type, $x^3$-independent solutions
 in the infinite space, and we have found that in this case the excitation
 has infinite energy because of the radiative component. Nevertheless, the
 obtained solution might be of physical interest. For example,
 in a real system like a superconductor of finite volume one could sustain
 the excitation by an external electromagnetic wave and the standing wave
 regime could be obtained in the whole volume of the superconductor. It is
 clear that  obtaining the standing wave in the infinite volume would
 require infinite energy.

  For description of the
 excitation created dynamically during, e.g., vortex reconnections,
 the standing wave, $x^3$-independent solution has to be modified.
 In such an event the excitation is created almost suddenly at
 certain  instant of time  with a finite  amount of energy
 and it occupies only a small part of the vortex.
 It turns out that it is not difficult to appropriately
 generalize our calculations, so that one can give approximate, analytic
 description of propagation of such a localised excitation along the vortex,
 as well as of propagation of radiation wave emitted from the travelling
 excitation \cite{20}.  Solutions of this type have finite energy.

\end{document}